\documentclass[journal=jacsat,manuscript=article]{achemso}

\usepackage{amsmath}
\usepackage{epstopdf}
\usepackage{graphicx}
\usepackage{dcolumn}
\usepackage{bm}
\usepackage{braket}
\usepackage{gensymb}
\usepackage{array}
\usepackage[utf8]{inputenc}
\usepackage[T1]{fontenc}
\usepackage{mathptmx}

\newcommand{\tinter}{$t_{\textrm{inter}}$}
\newcommand{\tintra}{$t_{\textrm{intra}}$}

\newcommand{\buplus}[1]{${#1}^1B_u^+$}
\newcommand{\buminus}[1]{${#1}^1B_u^-$}
\newcommand{\TTTsep}{$^3|T \cdots T\rangle$}
\newcommand{\QTTsep}{$^5|T \cdots T\rangle$}
\newcommand{\TTsep}{$^1|T \cdots T\rangle$}
\newcommand{\TT}[1]{$^1|TT\rangle_{#1}$}

\newcommand{\agminus}[1]{${#1}^1A_g^-$}

\title{Singlet Fission in Lycopene H-Aggregates}

\author{William Barford}
\affiliation{Department of Chemistry, Physical and Theoretical Chemistry Laboratory,\\ University of Oxford, Oxford, OX1 3QZ, United Kingdom}
\email{william.barford@chem.ox.ac.uk}

\begin{document}

\begin{abstract}
A  theory of singlet fission (SF)  in carotenoid dimers
is applied to explain the SF in lycopene H-aggregates observed after high energy photoexcitation.
The explanation proposed here is that a high energy, delocalized bright ($^1B_u^-$) state first relaxes and localizes onto a single lycopene monomer. The high-energy intramonomer state then undergoes internal conversion to  the  $1^1B_u^-$ state. Once populated, the $1^1B_u^-$ state allows exothermic bimolecular singlet fission, while its internal conversion to the $2^1A_g^-$ state is symmetry forbidden. The simulation of SF predicts that the intramonomer triplet-pair state undergoes complete population transfer to the intermonomer singlet triplet-pair state within 100 ps. ZFS interactions then begin to partially populate the  intermonomer quintet triplet-pair state up to ca.\ 2 ns, after which hyperfine interactions thermally equilibrate the  triplet-pair states, thus forming free, single triplets  within ca.\ 0.1 $\mu$s.
\end{abstract}

\vfill\pagebreak

Singlet fission in polyacenes is a photophysical process that has been studied for over 50 years,
with numerous reviews covering the topic\cite{Smith2010,Casanova2018,Musser19,Sanders19,Zhu2019}. In that time a consensus seems to have emerged as to the initial mechanisms of this process: namely, that after photoexcitation into the bright singlet state of a single chromophore, this state undergoes bimolecular fission into two triplets via a two-step electron-hole transfer\cite{Reichman2013b,Mazumdar2015,Casanova2018}. Research into singlet fission in polyacenes is now generally focussed on how to improve the triplet yield in pursuit of technological applications, which requires a full understanding of the fate of the triplet-pair\cite{Sanders19}.

In contrast, an understanding of singlet fission in oligoenes\cite{Rao2022,Rao2023}, polyenes\cite{Kraabel98,Lanzani99,Musser13} and carotenoids\cite{Tauber2010,Wang11,Zhu2014,Musser15,Llansola-Portoles2018,Dasgupta2021,Quaranta2021} is still in its infancy\cite{Zimmerman2018,Musser19,Ghosh2022,Manawadu2023a,Barford2023}. A consensus does not exist on the fundamental question as to whether singlet fission in these systems proceeds in the same manner as for polyacenes\cite{Musser15}, or whether an intermediate intramolecular correlated triplet-pair state is involved in the process\cite{Gradinaru2001,Rondonuwu2003}.

The possibility that an intermediate intramolecular triplet-pair state is involved in singlet fission is suggested by the well-known fact that the lowest excited singlet states of polyenes are indeed a superposition of correlated triplet-pairs\cite{Hudson72,Schulten72,Barford2022} and charge-transfer exciton states\cite{Barford2022} that are populated within 50 fs of photoexcitation of the `bright' Frenkel exciton state. However, since the triplets in the \emph{lowest} energy `dark' state (i.e., the $2^1A_g$ state) are  strongly bound\cite{Valentine2020,Barford2022}, a theory that involves \emph{a} `dark' state in singlet fission also has to explain why it is an exothermic process with a high yield of free triplets.

In this letter we develop a theoretical model of singlet fission that qualitatively explains a recent experimental investigation of singlet fission in lycopene H-aggregates by Kundu and Dasgupta\cite{Dasgupta2021}. In their work, Kundu and Dasgupta observed that singlet fission only occurs when the lycopene H-aggregates are excited at ca.\ 350 nm (ca.\ 3.5 eV), i.e., directly into the blue-shifted H-aggregate absorption band. In contrast, excitation in the 400 - 500 nm (2.5 - 3.1 eV) range, which corresponds to the monomer absorption band, does not cause singlet fission (see Figure 1 of ref\cite{Dasgupta2021}). This observation indicates that excess energy is needed to facilitate singlet fission.

\begin{figure}\label{Fi:1}
\includegraphics[width=0.5\textwidth]{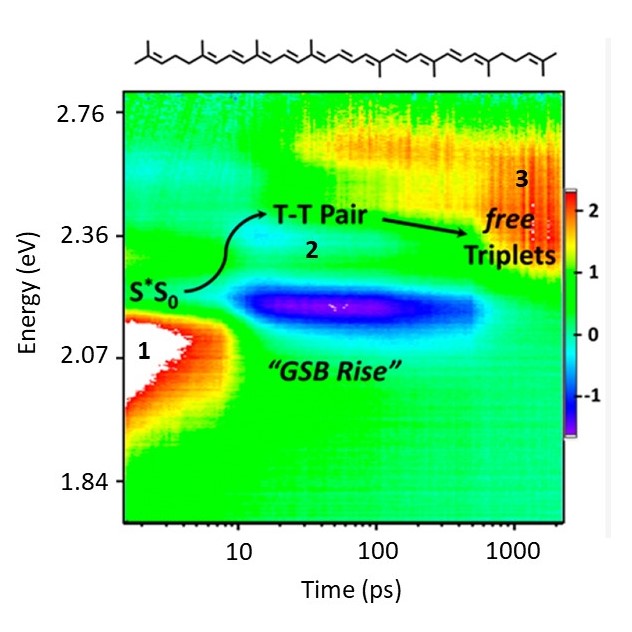}
\caption{The measured\cite{Dasgupta2021} transient excited-state-absorption of a lycopene H-aggregate following photoexcitation at 3.15 eV. See  Fig.\ 4 and the text following it for an explanation of the photophysics.
Reproduced and modified with permission from Kundu, A., Dasgupta, J., J.\  Phys.\ Chem.\ Lett.\ 2021, 1\textbf{2}, 1468.
}
\end{figure}

Another important observation of Kundu and Dasgupta is that the transient excited state absorption reveals an intermediate state generated after photoexcitation that has the characteristic signature of a member of the `$2A_g$' family of strongly correlated states. As has been shown by Barford and co-workers\cite{Valentine2020, Manawadu2023a}, it is the  charge-transfer exciton component of these correlated states that absorbs at ca.\ 2.1 eV, at an energy red-shifted by 0.3 eV from the free-triplet signal. This is precisely what is observed by Kundu and Dasgupta, as shown in Fig.\ 1. Importantly, the modeling of their transient data led  Kundu and Dasgupta to associate this feature as a higher-energy, intermediate triplet-pair state, i.e., \buminus{1}\ or $S_1^*$. This is significant, because as shown below, unlike for the \agminus{2}\ state, exothermic intermonomer singlet fission is possible from the \buminus{1}\ state\cite{Valentine2020,Manawadu2022}.

Using time-dependent DMRG calculations of the Hubbard-UV-Peierls model, Manawadu and co-workers\cite{Manawadu2022,Manawadu2023b} simulated the internal conversion from the bright state in the related carotenoid, zeaxanthin. Zeaxanthin has  18 conjugated C-atoms (i.e., 9 double bonds) and like lycopene (shown in Fig.\ 1) possesses $C_2$ symmetry. According to the simulation, excitation into the bright \buplus{1}\ state is followed within 10 fs by adiabatic internal conversion to the \buminus{1}\ state  via an avoided crossing of $S_2$ and $S_3$. However, as a consequence of $C_2$ symmetry, to zeroth-order in the Born-Oppenheimer approximation, subsequent internal conversion to  the \agminus{2}\ state ($S_1$) is  symmetry forbidden.

Here, we propose  a somewhat different  mechanism of internal conversion for a lycopene monomer within a H-aggregate. In particular, optical excitation into the blue-shifted absorption band of the H-aggregate excites a high-energy \buplus{}\  state that is delocalized over a number of monomers. As a consequence of electron-nuclear coupling and intermolecular interactions, this state  rapidly relaxes and localizes onto a single monomer. It then  undergoes nonadiabatic internal conversion  to an intramonomer \buminus{1}\ state. (This process is similar to the nonadiabatic relaxation and localization of high energy excited states of a conjugated polymer onto a single chromophore described in ref\cite{Mannouch2018}.) The timescale for internal conversion is determined by how fast energy is dissipated, but it is expected to ca.\ 100 fs\cite{Mannouch2018}.
Bimolecular exothermic singlet fission then follows, as the  relaxed energy of the \buminus{1}\ state lies ca.\ 0.4 eV above the  relaxed energies of a pair of triplets on separate monomers. Importantly, because of symmetry constraints, subsequent internal conversion from the  \buminus{1}\ state to the \agminus{2}\ state is a slow Herzberg-Teller-allowed process.
Conversely,  bimolecular interstate conversion from \buminus{1}\ to $T_1^i\otimes T_1^j$ (where $i$ and $j$ label the monomers) is not symmetry forbidden provided that the monomers in the H-aggregate are not perfectly aligned.
This scheme is shown schematically in Fig.\ 2.

Following the generation of intermonomer triplets,  the triplets diffuse into the aggregate, thus preventing intermonomer recombination into the \agminus{2}\ state.
Alternatively, as shown in ref\cite{Manawadu2023a}, additional torsional relaxation can stabilize the free triplets relative to the \agminus{2}\ state, thus making recombination an endothermic process.

\begin{figure}\label{Fi:2}
\includegraphics[width=0.5\textwidth]{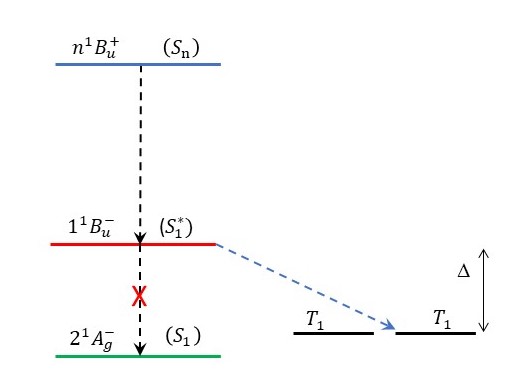}
\caption{A schematic energy level diagram for some states of lycopene (the approximate adiabatic state labels are in parentheses). The singlet states are on the left, while on the right is the energy level of two triplet states on separate monomers. 
Here we assume that there is localization and internal conversion from a high-energy, delocalized bright state (here labeled $S_n$) into the intramonomer \buminus{1}\ state, which then undergoes exothermic intermonomer singlet fission. To zeroth-order in the Born-Oppenheimer approximation, interconversion from the \buminus{1}\ state to the \agminus{2}\ state is symmetry forbidden. Bimolecular interstate conversion from \buminus{1}\ to $T_1^i\otimes T_1^j$ (where $i$ and $j$ label the monomers) is not symmetry forbidden provided that the monomers in the H-aggregate are not perfectly aligned.
$\Delta$ is the exothermic driving energy from the \buminus{1}\ state.
}
\end{figure}

\begin{figure}\label{Fi:3}
\includegraphics[width=1.0\textwidth]{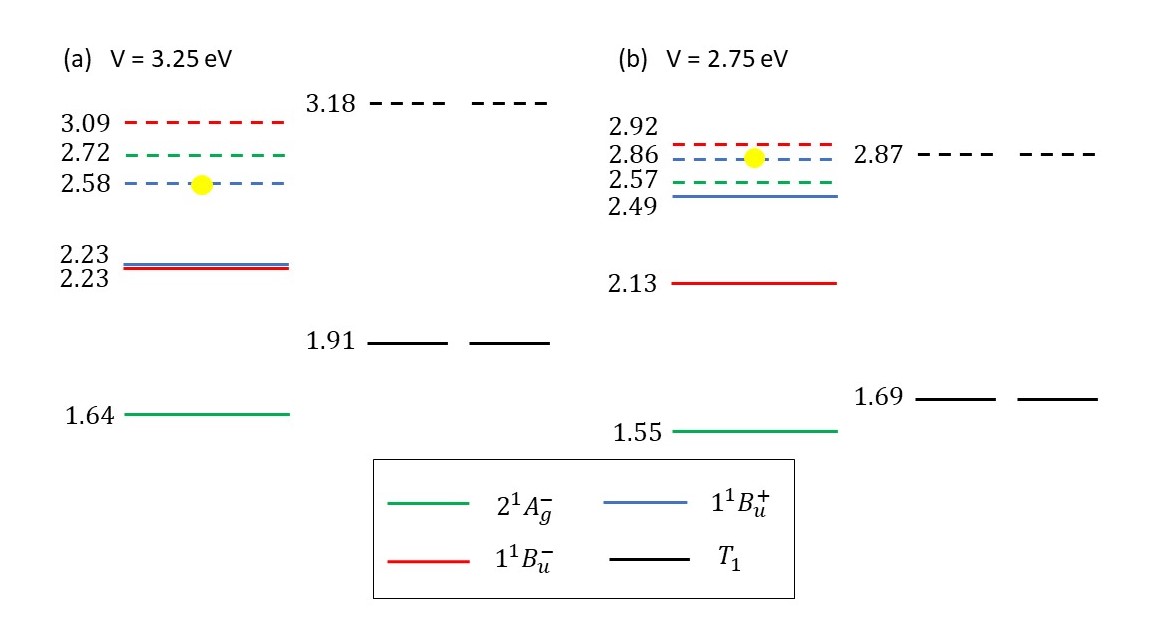}
\caption{Vertical (dashed lines) and relaxed (solid lines) excitation energies (in eV) of the low-energy states of monomeric lycopene. These are obtained by DMRG calculations of the Hubbard-UV-Peierls model, as explained in the SI.
The \emph{intra}monomer singlet states are on the left of each panel, while on the right of each panel is the energy level of two triplet states on separate monomers. In both cases intermonomer singlet fission is (potentially) exothermic (endothermic) from the \buminus{1}\ (\agminus{2}) state. The yellow circle indicates the initially excited monomeric `bright' state.  (a) The nearest-neighbor Coulomb repulsion, $V=3.25$ eV; in this case there is  level crossing between the \buplus{1}\ and \agminus{2}\ states.
(b) $V=2.75$ eV; in this case there is  level crossing between the \buplus{1}\ and \buminus{1}\ states.}
\end{figure}

Before discussing our simulations of the singlet fission process, we first address the question as to why excitation into the monomer absorption band does not cause singlet fission in lycopene\cite{Dasgupta2021}. To answer this question we turn to the  energy levels of monomeric lycopene obtained via DMRG calculations\cite{Barford01} of the Hubbard-UV-Peierls model for a monomer of 22 carbon sites. (Details of this model and its parametrization are given in the SI.)

In both the theoretical and experimental literature there is uncertainty about the exact relative energies of the vertical \buplus{1}, \agminus{2}, and \buminus{1}\ states in carotenoids.  These energies are highly sensitive to monomer-length and dielectric screening. Some high-level ab initio calculations\cite{Taffet19,Khokhlov20} predict that the vertical \agminus{2} state lies above the vertical \buplus{1}\ state. In this case, excitation into the   \buplus{1}\ state causes a \agminus{2}\ -- \buplus{1} level crossing. Alternatively, some semiemprically parametrized Hamiltonians\cite{Manawadu2022} predict that the vertical \agminus{2} state lies below the vertical \buplus{1}\ state, while the vertical \buminus{1}\ lies above it. In this case, excitation into  \buplus{1}\ causes a \buminus{1}\ -- \buplus{1} level crossing. In previous work, we have modeled both situations by modifying the parameters of the Hubbard-UV model\cite{Manawadu2023a}. The energy levels for lycopene for the two different semiempirical parameter sets are shown in Fig.\ 3. Panel (a) illustrates the \agminus{2}\ -- \buplus{1} level crossing while panel (b) illustrates the \buminus{1}\ -- \buplus{1} level crossing.

The absence of singlet-fission in lycopene under low-energy excitation is explained if a \agminus{2}\ -- \buplus{1} level crossing occurs (i.e., case (a)). In this case,  the \agminus{2}\ state is populated from the \buplus{1}\ state via Herzberg-Teller coupling, and
from which singlet fission is endothermic. However, high-energy excitation of a H-aggregrate implies that a bright \buplus{}\ state lies higher in energy than the \buminus{1}\ state, allowing the latter to be populated (as discussed above and shown schematically in Fig.\ 2). As already mentioned, once populated  the \buminus{1}\ state can undergo exothermic intermonomer singlet fission, or slow (Herzberg-Teller allowed) internal conversion to the \agminus{2} state.

Assuming, now,  that a singlet triplet-pair state has been formed (i.e., \buminus{1}), we  turn to discuss the singlet fission process. To simulate this process, we adopt the theory  introduced by Barford and Chambers to explain singlet fission in carotenoid dimers\cite{Barford2023}. The key assumptions of this theory are that singlet fission in carotenoid dimers occurs via one of the highly correlated dark states and that these dark states may be regarded as composed entirely of a singlet triplet-pair. Therefore, the charge-transfer exciton component of the dark state\cite{Barford2022} is assumed to be virtual component that acts to mediate the large intramonomer nearest neighbor triplet exchange interaction,\cite{Kollmar1993,Barford2023}
\begin{equation}\label{}
\hat{H}_{\textrm{exchange}} = {J}\hat{\textbf{S}}_i \cdot \hat{\textbf{S}}_{i+1}.
\end{equation}
Here, $\hat{\textbf{S}}$ is the  spin-1 (triplet) operator and $J$ is the \emph{inter}triplet exchange interaction. Thus, the intramonomer singlet and triplet triplet-pairs experience a nearest-neighbor attraction, $+2J$ and $ +J$, respectively, while the quintet triplet-pair experiences a nearest-neighbor repulsion $ -J$. As shown in ref\cite{Barford2023}, a value of $J = 1.23$ eV reproduces the intramonomer  $1^5A_g^-$ -- \agminus{2}\ energy gap of $0.4$ eV\cite{Valentine2020}.

Triplets hop between ethylene dimers on the same lycopene monomer with a transfer integral, \tintra\ = 0.88 eV\cite{Barford2023}, which results in a band of bound singlet triplet-pair states (the `$2A_g$' family of states\cite{Valentine2020}), i.e., \agminus{2}, \buminus{1}, \agminus{3}, $\cdots$. For convenience, these strongly interacting, intramonomer states are labeled \TT.
Higher in energy is a band of noninteracting intramonomer triplet-pair states.

Intermonomer triplet-pair states do not experience the strong intramonomer exchange interaction and consequently (neglecting for the moment the dipolar interaction to introduced shortly), the intermonomer triplet-pairs are noninteracting and space-separated. These pairs are labeled \TTsep, \TTTsep\ and \QTTsep\ for the singlet, triplet and quintet  states, respectively.

Triplets hop between adjacent ethylene dimers on neighboring lycopene monomers with a transfer integral, \tinter\ = 0.0088 eV. This mechanism causes the intra and inter monomer singlet triplet-pairs to hybridize to form the lowest energy bimolecular singlet state,
\begin{equation}\label{Eq:1140}
  ^1|\Psi\rangle = a ^1|TT\rangle     + b ^1|T \cdots T\rangle.
\end{equation}

As is shown in ref\cite{Barford2023}, the mixing ratio $a/b$ depends on the energy difference, $\Delta$, between \TT\ and \TTsep, with an exothermic process implying that $|b|^2 > |a|^2$. As also shown in  ref\cite{Barford2023}, a key emergent energy scale is $\Delta E_{QS}$, the exchange energy between $^1|\Psi\rangle$ and the intermonomer quintet state, \QTTsep. If $\Delta E_{QS} \ll k_BT$ full singlet fission occurs and the population of the initial \TT\ becomes equally equilibrated between the singlet, triplet and quintet intermonomer pairs, implying free triplets on separate monomers. In this simulation the exothermic driving energy (defined in Fig.\ 2) is taken to be $\Delta = 0.44$ eV. This value is large enough to ensure that $\Delta E_{QS}$ is small enough such that singlet fission goes to completion at long times. It is also consistent with the energy level diagram shown in Fig.\ 3. Using this value of $\Delta$, $\Delta E_{QS} = 1.16\times 10^{-3}$ eV.

The final interaction to include in the triplet-pair Hamiltonian is the \emph{intra}triplet dipolar (or zero-field-splitting) interaction. Assuming an axis of quantization along the principal dipolar axis, $Z$, this interaction reads\cite{Weil}
\begin{equation}\label{}
\hat{H}_{\textrm{ZFS}}^{\textrm{intra}} = \sum_{i=1}^2 D\left(\hat{S}^2_{iZ} - \frac{1}{3}\hat{S}_i^2 \right) +\frac{E}{2}\left(\hat{S}_{i+}^2 + \hat{S}_{i-}^2\right),
\end{equation}
where the sum is over both triplets in the pair, $\hat{S}$ are again the spin-1 operators, and $\hat{S}_{\pm}$ are the angular momentum shift operators. The first term couples the singlet triplet-pair with the $S_Z = 0$ component of the quintet triplet-pair, while the second term couples the singlet triplet-pair with the $S_Z = \pm 2$ components of the quintet triplet-pair. Since $D \sim 10^{-5}$ eV is already small compared to other energy scales and $E$ is typically 10 to 100 times smaller than D\cite{Weil,Kollmar1993}, the second term is neglected here.

In this work we take the  intermonomer triplet transfer integral, \tinter, to be a parameter. In particular, the choice of \tinter/\tintra\ = 0.01 predicts a \TT\ half-life of ca.\ 10 ps, which is consistent with experimental observations\cite{Dasgupta2021}. As shown in the SI, this choice implies a separation between the lycopene monomers in the H-aggregate of ca.\ 3 \AA, which is consistent with the experimental  absorption blue-shift of 0.92 eV.

We now turn to discuss the  dynamical simulation. The triplet-pair dynamics are determined by the quantum Liouville equation, which is computed in the eigenstate basis of the two-monomer triplet-pair Hamiltonian\cite{Barford2023}. Assuming the secular approximation, the quantum Liouville equation\cite{Nitzan2006,Kuhn2011}
for the populations $P_a \equiv \rho_{aa}$ is,
\begin{equation}\label{Eq:114}
  \frac{d P_{a}}{dt} = - \sum_{b \ne a}\left( k_{a  b} P_{a} - k_{b  a} P_{b} \right)
\end{equation}
while for the coherences it is,
\begin{equation}\label{Eq:115}
  \frac{d \rho_{ab}}{dt} = -i \omega_{ab} \rho_{ab} - 2\Gamma_{ab}(1-\delta_{ab})\rho_{ab}.
\end{equation}
The Bohr frequencies $\omega_{ab} = (E_a - E_b)/\hbar$, while $2\Gamma_{ab} = (\gamma_a + \gamma_b)$ and $\gamma_a = \sum_{b\ne a}  k_{a  b}$.

The inclusion of the ZFS interaction means that the energy eigenstates are not eigenstates of total spin. In this simulation we include both nonmagnetic and magnetic dephasing process. Thus, the thermal rates are a sum of the spin-conserving (SC) and spin-nonconserving (SNC) rates, i.e.,
  $k_{a  b} =  (k_{a  b}^{\textrm{SC}} +  k_{a  b}^{\textrm{SNC}})$,
where $k_{a  b}^{\textrm{SC}}$ and $k_{a  b}^{\textrm{SNC}}$ are defined in the SI.
Taking a reorganization energy of $0.05$ eV\cite{Reichman2013b}, at 300 K the nonmagnetic dephasing time is calculated to be ca.\ 1 ps\cite{Barford2023}. We take the characteristic time for the transverse ($S_z$-conserving magnetic) dephasing, $T_2$, to be 10 ns. Assuming only transverse-spin dephasing and neglecting the spin-flip component of $\hat{H}_{\textrm{ZFS}}^{\textrm{intra}}$ means that only the $S_Z = 0$ components of the triplet and quintet triplet-pairs states are connected to the singlet triplet-pairs states.

The triplet-pair basis and the two-monomer triplet-pair Hamiltonian are described in Section 2 of the SI; the parameters used in the simulation are listed in Section 3 of the SI. The solution of eqn (\ref{Eq:114}) and eqn (\ref{Eq:115}) is described in ref\cite{Barford2023}.

\begin{figure}\label{Fi:4}
\includegraphics[width=0.7\textwidth]{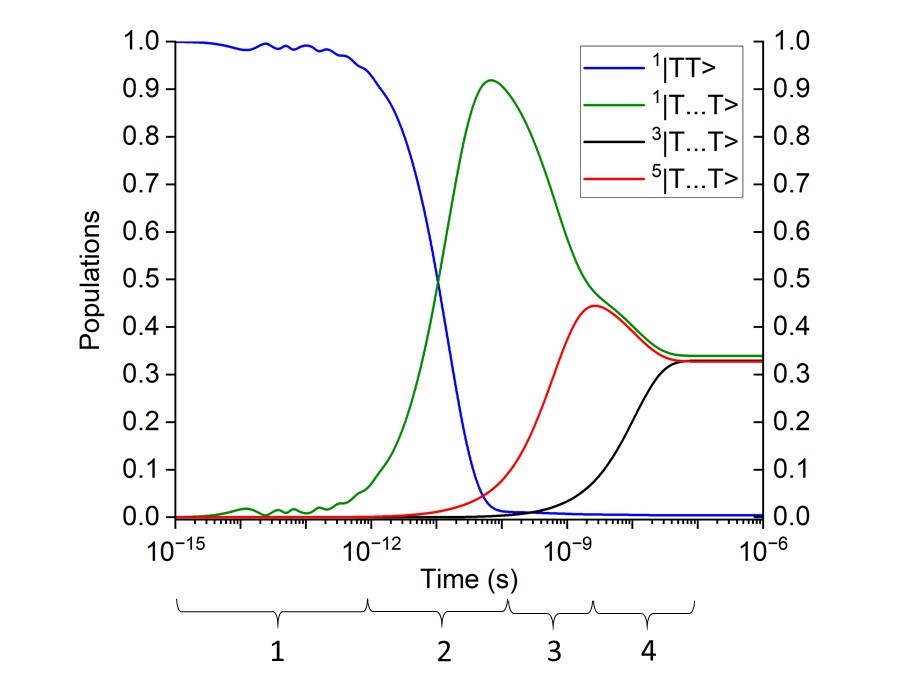}
\caption{The populations  as a function of time of the intramonomer singlet triplet-pair state, \TT{}, and the intermonomer singlet, triplet and quintet triplet-pair states, \TTsep, \TTTsep\ and \QTTsep. The four time regimes are discussed in the text. This figure should be compared to Fig.\ 1.
}
\end{figure}

We now discuss the results of the dynamical simulation. Taking as our initial state $|\Psi\rangle =$ \TT\ $\equiv |1^1B_u^-\rangle$, the triplet-pair populations
are illustrated in Fig.\ 4. From this we can identify four time regimes:
\begin{enumerate}
\item{Up to ca.\ 1 ps there is weak coherent dynamics between the intramonomer triplet pair, \TT, and the intermonomer triplet pair, \TTsep. As the system is off-resonance, the population is primarily \TT. This is represented as `1' in the experimental transient excited state absorption\cite{Dasgupta2021} shown in Fig.\ 1.}
\item{From ca.\ 1 ps to ca.\ 100 ps spin-conserving dephasing causes almost complete population transfer from \TT\ to \TTsep. In so far as \TTsep\ is a noninteracting triplet-pair, this
is `singlet-fission', although the system is still spin-correlated and the expectation value of $S^2 = 0$.}
\item{Next, from ca.\ 100 ps to ca.\ 2 ns the ZFS interaction  mixes the singlet and quintet intermonomer triplet-pair states. Since this process is over damped, there are no oscillations. Moreover, as $D/\Delta E_{QS} \simeq 0.01$, the population ratio of \TTsep\ to \QTTsep\ does not equal 1:2, as it would do if $D/\Delta E_{QS} \gg 1$. As \TTsep\ and \QTTsep\  are spectroscopically indistinguishable, the time regimes 2 and 3 are represented as `2' in Fig.\ 1.}
\item{Finally, after ca.\ 2 ns  and within ca.\ 100 ns transverse spin-dephasing equilibrates \TTsep, \TTTsep\ and \QTTsep\ to a population of 1/3 each. Equivalently, this population corresponds to spin-uncorrelated, single triplets\cite{Scholes2015,Barford2023}  on separate monomers and is represented by `3' in Fig.\ 1.}
\end{enumerate}

In summary, the proposed mechanism of singlet fission in lycopene H-aggregates  is the following. Optical excitation of the aggregate excites a  high-energy bright state ($S_n$) that is partially delocalized over the aggregate. This state rapidly relaxes and localizes onto a single lycopene monomer, populating an intermediate, singlet triplet-pair state. This is the \buminus{1}\ state, often labeled $S_1^*$. As explained in this letter, this state is the second member of the strongly correlated `$2A_g$' family of states. It is the strongly bound, intramonomer triplet-pair state, also labeled \TT{}, whose population dynamics are illustrated in Fig.\ 4.
Internal conversion from the \buminus{1}\ state to the \agminus{2}\ state is assumed to be slow, because it is symmetry forbidden.
Next, the \buminus{1}\ state undergoes `fission' into noninteracting, spin-correlated triplet pairs on separate monomers, labeled \TTsep. Because fission from the  \buminus{1}\ state is quite exothermic, within ca.\ 100 ps there is almost complete population transfer from \TT{}\ to \TTsep. Following the population of \TTsep, the ZFS interaction mixes \TTsep\ and \QTTsep. Finally, hyperfine interactions mix \TTsep, \TTTsep\ and \QTTsep. Since $\Delta E_{QS}/ k_BT \simeq 0.04$ (and we only consider transverse dephasing), the populations are equal. This final mixed state corresponds to spin-uncorrelated, single triplets on separate monomers\cite{Scholes2015,Barford2023}, i.e., $|\textrm{T}_1\rangle + |\textrm{T}_1\rangle$.

This letter has focussed on singlet fission in lycopene H-aggregates. However, other carotenoid H-aggregates also exhibit singlet fission with photophysical behavior semiquantitatively similar to that described here. For example, Quaranta and co-workers\cite{Quaranta2021} investigated singlet fission in lutein and violaxanthin H-aggregates. In common with lycopene, these carotenoids posses $C_2$ symmetry. Following photoexcitation of the aggregate, Quaranta and co-workers\cite{Quaranta2021} propose an intermediate state that participates in singlet fission, which they nominated as a vibrationally hot $S_1$ state. In light of the work described here, however, we think that this state is more likely to be a distinct (albeit a related) electronic state, namely the \buminus{1}\ state.

In conclusion, this letter describes a theory of singlet fission in lycopene H-aggregates that is in semiquantitative agreement with experimental observations\cite{Dasgupta2021}. In particular, the theory assumes that singlet fission in carotenoid systems occurs via an intermediate,  intramonomer singlet triplet-pair state (i.e., \buminus{1}), which facilitates exothermic intermonomer singlet fission. This state is populated via the excitation of a higher energy H-aggregate bright state. In contrast, singlet fission in polyacenes  occurs directly from the intramolecular bright state. Thus, the participation of an intramolecular triplet-pair state  in carotenoid singlet fission implies that this mechanism is quite different from that of polyacenes.

\vfill
\pagebreak



\section{Supporting Information}

\section{1.\ The Hubbard-UV-Peierls Hamiltonian}\label{Se:1}

The DMRG calculations\cite{Barford01,Manawadu2023b} of the electronic states of lycopene were performed using the Hubbard-UV-Peierls Hamiltonian. This Hamiltonian has three components.

The purely  electronic Hubbard-UV Hamiltonian is
\begin{equation}
\hat{H}_{\textrm{UV}} = -2\sum_{n=1}^{N-1} \beta_n \hat{T}_n + U \sum_{n=1}^N  \big( \hat{N}_{n \uparrow} - \frac{1}{2} \big) \big( \hat{N}_{n \downarrow} - \frac{1}{2} \big)
+ \frac{1}{2} \sum_{n=1}^{N-1} V \big( \hat{N}_n -1 \big) \big( \hat{N}_{n+1} -1 \big),
\end{equation}
which contains a nearest neighbor electron transfer term, $\beta_n$, and onsite and nearest neighbor Coulomb interactions, $U$ and $V$, respectively.
$\hat{T}_ n = \frac{1}{2}\sum_{\sigma} \left( c^{\dag}_{n , \sigma} c_{n + 1 , \sigma} + c^{\dag}_{n+ 1 , \sigma} c_{n  , \sigma} \right)$ is the bond order operator, $\hat{N}_n$ is the number operator and $N$ ($=22$ for lycopene) is the number of conjugated carbon-atoms ($N/2$ is the number of double bonds).

The electrons couple to the nuclei via changes in the C-C bond length (which changes the effective electron transfer integral)
\begin{equation}
	\hat{H}_{\textrm{e-n}} =  2 \alpha \sum_{n=1}^{N-1}  \left( u_{n+1} - u_n \right) \hat{T}_n ,
\end{equation}
where $\alpha$ is the electron-nuclear coupling parameter and $u_n$ is the displacement of nucleus $n$ from its undistorted position.

Finally, the nuclear potential
energy is described by
\begin{equation}
	\hat{H}_{\textrm{elastic}} = \frac{K}{2} \sum_{n=1}^{N-1} \left( u_{n+1} - u_n \right)^2,
\end{equation}
where
$K$ is the nuclear spring constant.

The Hubbard-UV-Peierls Hamiltonian is  defined as
\begin{equation}\label{Eq:UV-P}
\hat{H}_{\textrm{UVP}} = \hat{H}_{\textrm{UV}} + \hat{H}_{\textrm{e-n}} +	\hat{H}_{\textrm{elastic}}.
\end{equation}
This Hamiltonian  is invariant under both a two-fold proper rotation (i.e., a $C_{2}$ operation) and a particle-hole transformation (i.e., $(\hat{N}-1) \rightarrow -(\hat{N}-1)$), and so its eigenstates are labeled either  $A_g^{\pm}$ or $B_u^{\pm}$.

The parameters are the same as used in previous work\cite{Manawadu2022, Manawadu2023a}, i.e., $\beta= 2.4$ eV,  $U = 7.25$ eV,  $K = 46$ eV $\AA^{-2}$ and $\alpha = 4.6$ eV $\AA^{-1}$. As described in the main paper, in order to model relative \agminus{2}/\buplus{1}\ energies  we make two choices for $V$, i.e., $V = 2.75$ eV or $V = 3.25$ eV.

\section{2.\ The Triplet-Pair Basis and Hamiltonian}\label{Se:4}

The initial singlet triplet-pair state, \TT{}, is\cite{Barford2022}
\begin{equation}\label{Eq:01}
 {^1|\textrm{TT}\rangle} = \sum_{ij \in \times = 1} \Phi_{ij} {^1|i,j\rangle},
\end{equation}
where the singlet triplet-pair basis state is
\begin{equation}\label{Eq:02}
  ^1|i,j\rangle = \frac{1}{\sqrt{3}}\left( |1;i\rangle|-1;j\rangle - |0;i\rangle|0;j\rangle + |-1;i\rangle|1;j\rangle \right),
\end{equation}
$|S_Z;i\rangle$ is a triplet on ethylene dimer $i$ with spin projection $S_Z$, $\Phi_{ij}$ is the lowest eigenstate of the one-monomer Hamiltonian (defined below) with $B_u$ symmetry, and $ij \in \times = 1$ in eqn (\ref{Eq:01}) implies that dimers $i$ and $j$ are on monomer 1.

As described in the main paper, triplets on the same monomer experience the exchange interaction,\cite{Kollmar1993,Barford2023}
\begin{equation}\label{}
\hat{H}_{\textrm{exchange}} = {J}\hat{\textbf{S}}_i \cdot \hat{\textbf{S}}_{i+1},
\end{equation}
where $\hat{\textbf{S}}$ is the  spin-1 (triplet) operator and $J$ is the \emph{inter}triplet exchange interaction.

Triplets also experience the \emph{intra}triplet dipolar (or zero-field-splitting) interaction,
\begin{equation}\label{}
\hat{H}_{\textrm{ZFS}}^{\textrm{intra}} = \sum_{i=1}^2 D\left(\hat{S}^2_{iZ} - \frac{1}{3}\hat{S}_i^2 \right),
\end{equation}
where the only the $S_Z$ conserving component is retained.

In this work we investigate the role of transverse spin-dephasing and the component of the ZFS-Hamiltonian which connects the $S_z=0$ components of each total spin.
The $S_z = 0$ components of the triplet and quintet triplet-pair bases are,
\begin{equation}\label{}
  ^3|i,j\rangle = \frac{1}{\sqrt{2}}\left( |1;i\rangle|-1;j\rangle - |-1;i\rangle|1;j\rangle \right),
\end{equation}
and
\begin{equation}\label{}
  ^5|i,j\rangle = \frac{1}{\sqrt{6}}\left( |1;i\rangle|-1;j\rangle + 2|0;i\rangle|0;j\rangle + |-1;i\rangle|1;j\rangle \right),
\end{equation}
respectively.

\begin{figure}
\includegraphics[width=0.9\linewidth]{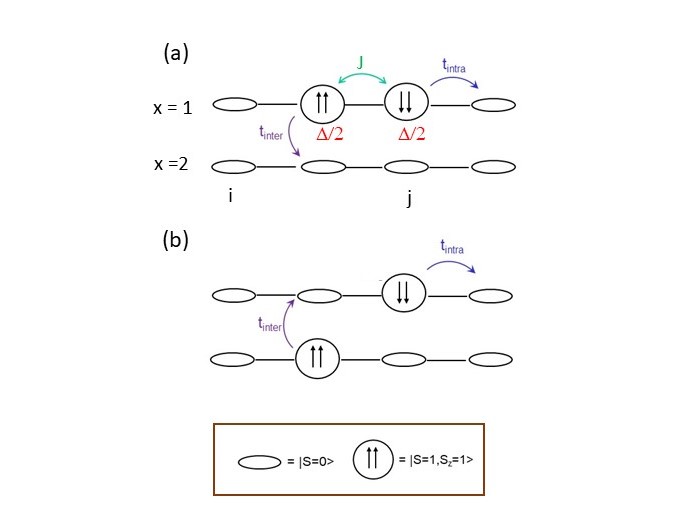}
\caption{A schematic illustration of a carotenoid dimer. (a) with a triplet-pair on monomer $\times=1$, with triplets on ethylene dimers $i$ and $j$.
Monomer $\times=2$ is in its ground state. This represents $\hat{H}_{\textrm{exchange}} + \hat{H}_{\textrm{single}}^{\times=1,2}$.
(b) represents $ \hat{H}_{\textrm{double}}+ \hat{H}_{\textrm{inter}}$.
\tintra\ and \tinter\ are the hopping matrix elements between neighboring intramonomer and intermonomer ethylene dimers, respectively.}
\label{Fi:18}
\end{figure}

The triplets hop between neighboring dimers on the same monomer (see Fig.\ 5(a)), described by
\begin{eqnarray}\label{Eq:401}
  \hat{H}_{\textrm{single}}^{\times=1,2} =
\Delta\sum_{i, j>i \in \times}  \left|i,j\rangle \langle i,j\right|
 +  t_{\textrm{intra}}\sum_{i \ne j \in \times}\left(\left|i \pm 1,j\rangle \langle i,j\right| + \textrm{H.C.} \right),
\end{eqnarray}
$\Delta$ is the exothermic driving energy of \buminus{1}\ relative to two single triplets on separate monomers.
Triplets on separate monomers hop between neighboring dimers on the same monomer (see Fig.\ 5(b)), described by
\begin{eqnarray}\label{Eq:402}
  \hat{H}_{\textrm{double}} =
t_{\textrm{intra}}\sum_{i \in \times=1} \sum_{j \in \times=2} \left[ \left(\left|i \pm 1,j\rangle \langle i,j\right| + \textrm{H.C.} \right)+
   \left(\left|i,j \pm 1\rangle \langle i,j\right| + \textrm{H.C.} \right) \right],
\end{eqnarray}
and between adjacent dimers on neighboring monomers (see Fig.\ 5(b)), described by
\begin{eqnarray}\label{Eq:403}
  \hat{H}_{\textrm{inter}} =
  t_{\textrm{inter}} \sum_{\times = 1}^2\sum_{i_{\times}} \sum_{j_{\times}>i_{\times}}
\left[
\left(
\left|i_{\times},j_{\times}\rangle \langle i_{\bar{\times}},j_{{\times}}\right|
+ \textrm{H.C.}
\right)  +
\left(
\left|i_{\times},j_{\times}\rangle \langle i_{{\times}},j_{\bar{\times}}\right|
+ \textrm{H.C.}
\right)
\right].
\end{eqnarray}


\section{3.\ Model Parameters}\label{Se:5}

Table 1 lists the parameters used in the two-monomer triplet-pair Hamiltonian described in Section 2, as well as the parameters used in the quantum Liouville equation.

\begin{table}[h]
\small\centering
{\renewcommand{\arraystretch}{1.2}
\begin{tabular}{|p{10cm}|p{3cm}|}
\hline
Parameter &  Value \\
\hline

Intramonomer triplet exchange interaction\cite{Barford2023}, $J$ & 1.23 eV \\
Intramonomer triplet transfer integral\cite{Barford2022}, \tintra & 0.88 eV \\
Intermonomer triplet transfer integral, \tinter\ & 0.0088 eV \\
Exothermic driving energy, $\Delta$ & 0.44 eV \\
Intratriplet dipolar interaction, $D$ & $10^{-5}$ eV \\
Reorganization energy, $\lambda$ & 0.05 eV \\
Spectral function cut-off frequency, $\omega_0$ &  0.2 eV \\
Temperature & 300 K\\
Spin-dephasing factor, $\gamma$ & $10^{-4}$ \\
Derived two-monomer quintet-singlet exchange energy, $\Delta E_{QS}$ & $1.16\times 10^{-3}$ eV\\
\hline
\end{tabular}}
\caption{Values of input and derived parameters.}
\label{Ta:2}
\end{table}

\section{4.\ Computation of Interstate Rates}\label{Se:2}

The inclusion of the ZFS interaction into the two-monomer triplet-pair Hamiltonian means that the energy eigenstates are not eigenstates of total spin. In the simulation we include both nonmagnetic and magnetic dephasing processes. To account for this
the thermal rates that appear in quantum Liouville equation are defined as the sum of the spin-conserving (SC) and spin-nonconserving (SNC) rates, i.e.,
\begin{equation}
  k_{a  b} =  k_{a  b}^{\textrm{SC}} +  k_{a  b}^{\textrm{SNC}}.
\end{equation}

Defining the Bohr frequencies as $\omega_{ab} = (E_a - E_b)/\hbar$ and taking $\omega_{ab} \geq 0$, the spin-conserving thermal rates are,\cite{Nitzan2006,Kuhn2011}
\begin{equation}\label{Eq:106}
  k_{a  b}^{\textrm{SC}} = \left(\frac{2 \lambda}{\hbar} \right) J(\omega_{ab})(n(\omega_{ab}) + 1) C_{ab}^{\textrm{SC}}
\end{equation}
and
\begin{equation}\label{Eq:107}
  k_{ b a}^{\textrm{SC}} = \left(\frac{2 \lambda}{\hbar} \right) J(\omega_{ab})n(\omega_{ab}) C_{ab}^{\textrm{SC}},
\end{equation}
where $n(\omega) = (\exp \beta\hbar\omega -1)^{-1}$ is the Bose distribution function, $J(\omega) = \omega \omega_0/(\omega^2+\omega_0^2)$ is the (dimensionless) Debye-spectral function, $\omega_0$ is the cut-off frequency and $\lambda$ is the bath reorganization energy. The parameters are listed in Table 1.

The spin-conserving overlap factors are,
\begin{equation}
C_{ab}^{\textrm{SC}} = 2\sum_{m} S_{ma}^2 S_{mb}^2.
\end{equation}
Here, $a$ and $b$ label energy eigenstates of the two-monomer Hamiltonian, whereas $m$ labels a real-space basis state of the triplet-pair states\cite{Barford2023}. In particular, $m$ encodes the dimer locations of each triplet in the pair, as well as the total spin of the pair.
$\textbf{S}$ is the matrix whose columns are the eigenvectors of the two-monomer Hamiltonian  represented in the real-space basis. Thus, $S_{ma}^2 = P_{ma}$ is the probability that the $a$th energy eigenstate occupies the $m$th real-space basis state.

Similarly, the spin-nonconserving thermal rates are,
\begin{equation}
  k_{a  b}^{\textrm{SNC}} = \gamma\left(\frac{2 \lambda}{\hbar} \right) J(\omega_{ab})(n(\omega_{ab}) + 1) C_{ab}^{\textrm{SNC}}
\end{equation}
and
\begin{equation}
  k_{ b a}^{\textrm{SNC}} = \gamma\left(\frac{2 \lambda}{\hbar} \right) J(\omega_{ab})n(\omega_{ab}) C_{ab}^{\textrm{SNC}},
\end{equation}
where $\gamma$ is a factor to take into account weaker transverse spin-dephasing than spin-conserving dephasing. We take $\gamma = 10^{-4}$, which at 300 K implies $T_2 \sim 10$ ns.

The spin-nonconserving overlap factors are,
\begin{equation}
C_{ab}^{\textrm{SNC}} = \sum_{m} f_{m\bar{m}}\left(S_{ma}^2 S_{\bar{m}b}^2 + S_{\bar{m}a}^2 S_{{m}b}^2\right),
\end{equation}
where the label $\bar{m}$ refers to  the same triplet-pair dimers as $m$, but corresponds to a different spin-eigenstate, i.e., singlet, triplet or quintet.
In addition, $f_{m\bar{m}} = 2/3$ for singlet-triplet transitions, $f_{m\bar{m}} = 1/3$ for triplet-quintet transitions, and $f_{m\bar{m}} = 0$ for singlet-quintet transitions\cite{Barford2023}.


\section{5.\ Intermonomer Triplet-Pair Coupling}\label{Se:3}

The simulations of singlet fission in this work took the intermonomer triplet transfer integral, \tinter, to be a parameter. In particular, the choice of \tinter/\tintra = 0.01 results in a half-life of the intramonomer singlet triplet-pair, \TT{}, to be ca.\ 10 ps. This value is consistent with the experimental observations of Kundu and Dasgupta\cite{Dasgupta2021} (see Fig.\ 1). Indeed, the  \TT{}\ half-life is rather sensitive to \tinter/\tintra: values of \tinter/\tintra = 0.01, 0.001 and 0.0001 predict half-lives of 10 ps, 1 ns and 100 ns, respectively.

As explained in ref\cite{Barford2023}, both intra and inter monomer triplet transfer is assumed to be a superexchange process, mediated by a virtual charge-transfer exciton. Thus, $t \propto \beta^2$, where $\beta$ is the  p$_z$-orbital resonance integral. We therefore deduce that $\beta_{\textrm{inter}}/\beta_{\textrm{intra}} =0.1$. Using the Mulliken expression\cite{Mulliken} for $\beta$, i.e.,
\begin{equation}\label{Eq:2}
  \beta = (10.6 \textrm{ eV})\times\exp(-r\xi)\left(1 + r\xi + \frac{2}{5}(r\xi)^2+ \frac{1}{15}(r\xi)^3 \right),
\end{equation}
where $r$ is in \AA\ and $\xi = 3.07$ \AA$^{-1}$, we can now estimate the intermolecular separation. Taking the single bond length $r_{\textrm{single}} = 1.45$ \AA\ implies that $\beta_{\textrm{intra}} = 2.4$ eV and thus $\beta_{\textrm{inter}} = 0.24$ eV. Again, using eqn (\ref{Eq:2}), we find that the intermonomer separation is thus predicted to be 2.9 \AA.

As we now show, a separation of ca.\ 3 \AA\ between the lycopene monomers in the H-aggregate is consistent with a blue shift of 0.92 eV, as measured in ref\cite{Dasgupta2021}. According to the line-dipole theory of exciton transfer integrals\cite{Barford07}, the exciton transfer integral, $J$, between two conjugated molecules of length $L$ and separation $r$ is
\begin{equation}\label{Eq:3}
 J = \left(\frac{\mu^2}{4\pi\epsilon_0 r^3}\right)\left(\frac{2}{(L/r)^2}\right)\left(1-\frac{1}{\sqrt{1+(L/r)^2}}\right),
\end{equation}
where $\mu$ is the transition dipole moment.

According to DMRG calculations\cite{Barford01}, $\mu = 5.71\times 10^{-29}$ Cm for a lycopene monomer of 22 conjugated C-atoms. The measured blue-shift in a H-aggregate is approximately $zJ$, where $z$ is the number of nearest neighbors to which each monomer is dipole coupled. Using $zJ = 0.92$ eV and $r = 2.9$ \AA, we find that $z = 4.6$. This result seems very reasonable and places confidence that, as well as being empirically justified, our parameter choice of \tinter/\tintra = 0.01 is physically realistic.


\vfill
\pagebreak

\bibliography{references}

\end{document}